\documentclass[a4paper]{spie}  

\usepackage[]{graphicx}
\usepackage{here}

\title{Design and manufacture of micro-optical arrays using 3D diamond machining techniques} 

\author{J\"urgen Schmoll\supit{a}, David J. Robertson\supit{a}, David A. Ryder\supit{a}
\skiplinehalf
\supit{a} Centre for Advanced Instrumentation, Netpark Research Institute, Joseph Swan Road, Netpark, Sedgefield TS21 3FB, United Kingdom \\
}

\authorinfo{Further author information: (Send correspondence to J.S.) \\
J.S.: E-mail: jurgen.schmoll@durham.ac.uk, Telephone: 0044-(0)191-334-4810 \\ }

\begin{document} 
  \maketitle 

\begin{abstract}
We describe our work towards the manufacture of micro-optical arrays using freeform diamond machining techniques. Simulations have been done to show the feasibility of manufacturing micro-lens arrays using the slow-tool servo method. Using this technique, master shapes can be produced for replication of micro-lens arrays of either epoxy-on-glass or monolthic glass types. A machine tool path programme has been developed on the machine software platform DIFFSYS, allowing the production of spherical, aspherical and toric arrays. In addition, in theory spatially varying lenslets, sparse arrays and dithered lenslet arrays (for high contrast applications) are possible to produce. In practice, due to the diamond tool limitations not all formats are feasible. Investigations into solving this problem have been carried out and a solution is presented here.
\footnote{Copyright 2006 Society of Photo-Optical Engineers. This paper will be published in SPIE Conf. Series 6273 and is made available as an electronic preprint with permission of SPIE. One print or electronic copy may be made for personal use only. Systematic or multiple reproduction, distribution to multiple locations via electronic or other means, duplication of any material in this paper for a fee or commercial purposes, or modification of the content of the paper are prohibited.}

\end{abstract}
\enlargethispage{10mm}


\section{INTRODUCTION}
The demands of modern astronomical micro-optical components in terms of surface shape, number of elements and packing density are increasingly more difficult to meet with classical optics manufacture. Apart from the difficulty of producing a large number of identical spherical or possibly aspheric surfaces, the alignment of the elements onto a common carrier is a difficult task. Furthermore the stability of this alignment, e.g. in cryogenic situations, is critical. Alternatively, machining of optical components out of a single block, creating the array or a negative array replication master, becomes increasingly popular especially in context of free-form machines. Once the numerical model is derived, the part can be CNC machined without interaction e.g. for alignment needs between the different surfaces. The CfAI has been involved in design and manufacture of different mirror arrays used for image-slicing integral-field-units (IFUs), based on the advanced image slicer design (Content 1998, \cite{Content1998}). Those arrays incorporate apheric and toric surfaces, that are achievable with 3D-machining techniques. However, experience with image-slicing arrays has shown that monolithic manufacture is not always possible due to tooling constraints. A similar constraint is described in context of a dithered lenslet array master. 
\label{sect:intro}  


\section{Micro lens arrays}
Micro-lens arrays are necessary for fiber-lenslet coupled IFUs. They adapt the incoming light cone to the fiber, while optimizing the fill-factor of the input array. At the output linear arrays are required to match the fast light cone emerging from the fiber towards a slower collimator. Instruments built in Durham employing this technique are SMIRFS (Haynes, 1998 \cite{Haynes1998}), TEIFU (Murray et al, 2000 \cite{Murray2000}), IFUs for GMOS (Allington-Smith et al, 2002 \cite{Allington-Smith2002}) and IMACS (Schmoll et al, 2004 \cite{Schmoll2004}). Other IFUs requiring microlenses, not necessary in conjunction with fibers, have been studied in context with million-element IFUs (Content et al, 2003 \cite{Content2003}). Those lenslet shapes can be arbitrary in terms of format and optical surface. Although in the fiber based IFUs produced by the CfAI concentric, hexagonally shaped lenslets were used, the million-element IFU described by Content (\cite{Content2003}) requires cylindrical, elongated lenslets. While for the existing IFUs mentioned commercially available microlens arrays have been used, it is planned to produce masters for epoxy-on-glass and monolithic glass arrays using our Nanotech 350 FG 5-axis UHP machine. In particular the slow tool servo method, as it was already used for micro lens array masters (Yi and Li, 2005 \cite{Yi2005}), is recognized as bearing potential for this process.

\subsection{Innovative microlens array designs}
Through the ability to produce in-house freeform lenslet array masters, the range of available arrays increases. New types of lenslet arrays are possible - dense or sparse arrays of round, square, hexagonal or totally arbitrarily shaped lenses, with optical surfaces being spherical, aspherical or toroidal. Two ideas for new lenslet array types are the foveal sampling and the dithered lenslet arrays.
\subsubsection{Foveal sampling lenslet arrays}
\begin{figure}[H]
   \begin{center}
   \begin{tabular}{c}
   \includegraphics[width=10cm]{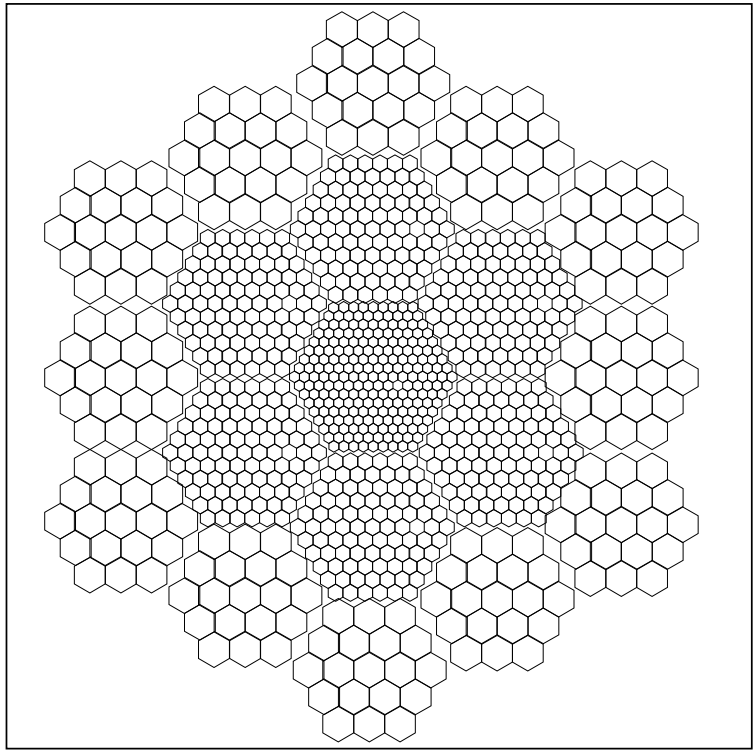}
   \end{tabular}
   \end{center}
   \caption{ \label{fig:tunnelvision_notext} 
Example of hexagonal lenslet array composite with three different spaxel sizes.}
\end{figure}

The sampling element size in an integral field unit has an important role for the spatial resolution of the data cube as well as for the signal/noise ratio of the spectrum. On objects with brighter and dimmer parts, the choice of the sampling element size is often a tradeoff between both, as smaller spaxels yield a higher spatial resolution but diminish the signal/noise ratio. When observing objects with a comparably bright center, the exposure may not yield the faint information to characterize the peripheral parts of the objects or the sky background. An IFU with variable sampling as depicted in fig. \ref{fig:tunnelvision_notext} could be a technical solution to this problem. To fully exploit such an IFU, different spectrographs will be necessary for each spaxel size. Such an IFU would get a high resolution at the center, e.g. on a galaxy core. An annulus around this bearing larger elements covers a larger array of sky and allows a better signal/noise ratio in areas where the object is dimmer. A third annulus with even coarser sampling around this will be useful to map areas of very low surface brightness, and/or to subtract the background sky from the object. Another use of this foveal technique can be found in non-multi conjugated adaptive optics systems where the image quality degrades with distance from the center due to the effects of leaving the isoplanatic region of correction. Here the spaxel size would adapt to the available spot size, with the telescope system being diffraction limited only in the central area.

\subsubsection{Dithered lenslet arrays}
\begin{figure}[H]
   \begin{center}
   \begin{tabular}{c}
   \includegraphics[width=10cm]{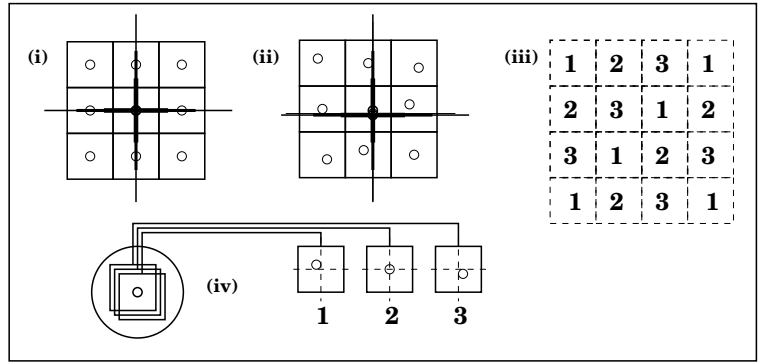}
   \end{tabular}
   \end{center}
   \caption{ \label{fig:larr_dither} 
Dithered lenslet array principle: (i) Non-dithered array, (ii) dithered array, (iii) method of paving an array with 3 different lenslet elements as depicted in (iv).}
\end{figure}

A well-known problem of existing micro-lens arrays is the crosstalk between adjacent apertures, diminishing the contrast especially in situations where a high dynamic range is to be observed, as dim objects around bright sources. As shown at by Roth et al \cite{Roth2005}, particular spikes are created at each microlens focus, running straight through the foci of adjacent lenslets. Those spikes are caused by diffraction and border irregularities between adjacent lenslets, as a soft edge instead of a well-defined change of surface tilt where adjacent elements touch each other. As smaller the lenslet is, as more likely it becomes that a portion of light will hit adjacent foci, as shown in case (i) of figure \ref{fig:larr_dither}. A way to avoid this situation is a translation of the adjacent lens vertices as shown in (ii). The tiling method shown in (iii) and (iv) assures that in each direction the next and overnext neighbour focus of any particular lens will be missed by it's diffraction spike. The solution works similarly for hexagonal lenses, and three different lens types are necessary. One of it is concentric, while the other two have opposite offsets along the lens diagonal. The third neighbour focus will be hit by the spike again, but the light will arrive there in a very large angle of incidence. Hence it can be baffled out, or (in case of fibers) it will be filtered out by the numerical aperture limit of the fiber.

\subsection{Simulation of preform manufacture with the slow-tool servo}
Preparing the manufacture of freeform surfaces, software has been developed that can simulate the slow-tool servo mode. Initially running under IDL \footnote{IDL by Research Systems Inc., Boulder, Coloradu, USA}, the code later has been transferred to DIFFSYS. The lenslet sag depends on the distance from the vertex of a particular element and is specified by equation \ref{eq:sagequation}, with c being curvature, k the conic constant and $A_n$ the n-th aspheric coefficent.

\begin{equation}
\label{eq:sagequation}
Z = \frac{c r^2}{1+\sqrt{1-(1+k)c^2 r^2}} + \sum_{i=4,6,8,...} a_i r^{i}
\end{equation}

The software allows the simulation of square, rectangular and hexagonal arrays. Torodial shapes are also possible, and the vertex can be offset from the lenslet center. Using a data matrix, different lenslets types can be arranged into a common array.

\subsection{Dithered lenslet arrays as test case}

The manufacture of a dithered lenslet array preform using square lenses of 500 $\mu$m side length has been simulated to predict the implications of off-axis-optics towards the image quality. Apart from the monolithic approach, other solutions which may be more feasible will be discussed in the following. Each of the arrays in the following have a 400 $\mu$m pitch, focal ratios of f/4 and dither offsets of 50 $\mu$m. The manufacture scenario consists of diamond-machining a negative shape that is used as a master to press PMMA lenslets onto BK7 substrates. The simulation (as all others in this publication) is made in five wavelengths (400 to 800 nm), and a f/100 input beam leads to a maximum field angle of 0.3 degree. \\

\begin{figure}[H]
   \begin{center}
   \begin{tabular}{c}
   \includegraphics[width=10cm]{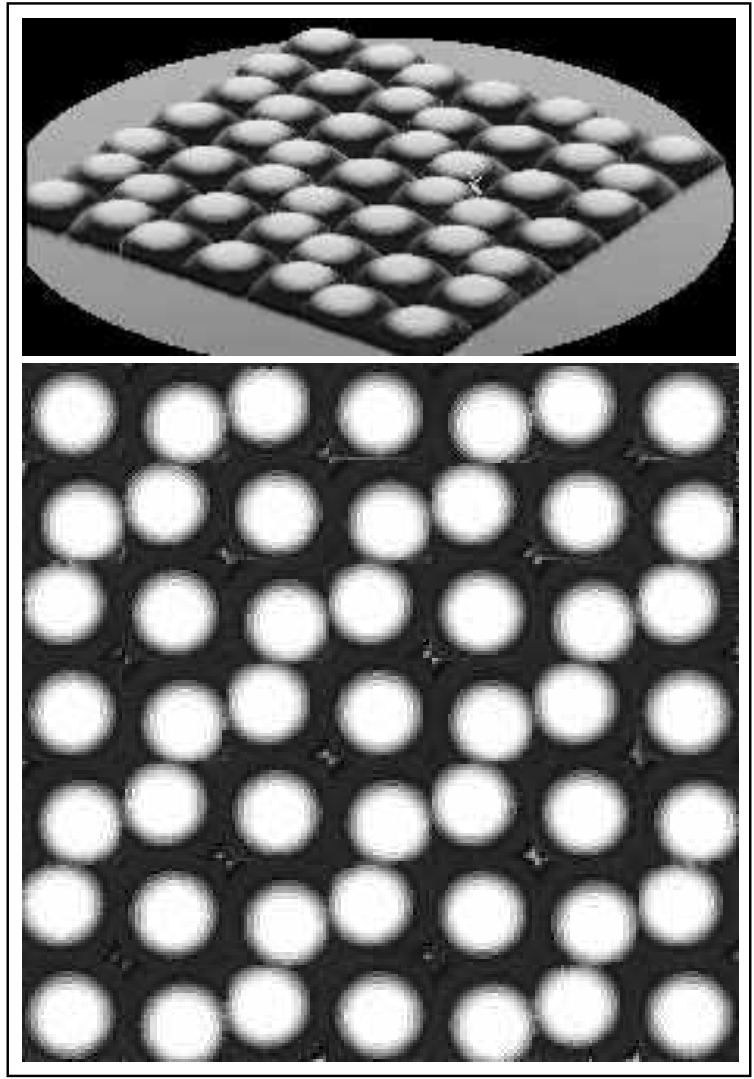}
   \end{tabular}
   \end{center}
   \caption{ \label{fig:ditherlarr7x7}
Modelling a monolithic 7x7 dithered lenslet array. The plot of the complete, spiral shaped tool path reveals the surface profile.}
\end{figure}


\subsubsection{Monolithic approach}
\begin{figure}[H]
   \begin{center}
   \begin{tabular}{c}
   \includegraphics[width=12cm]{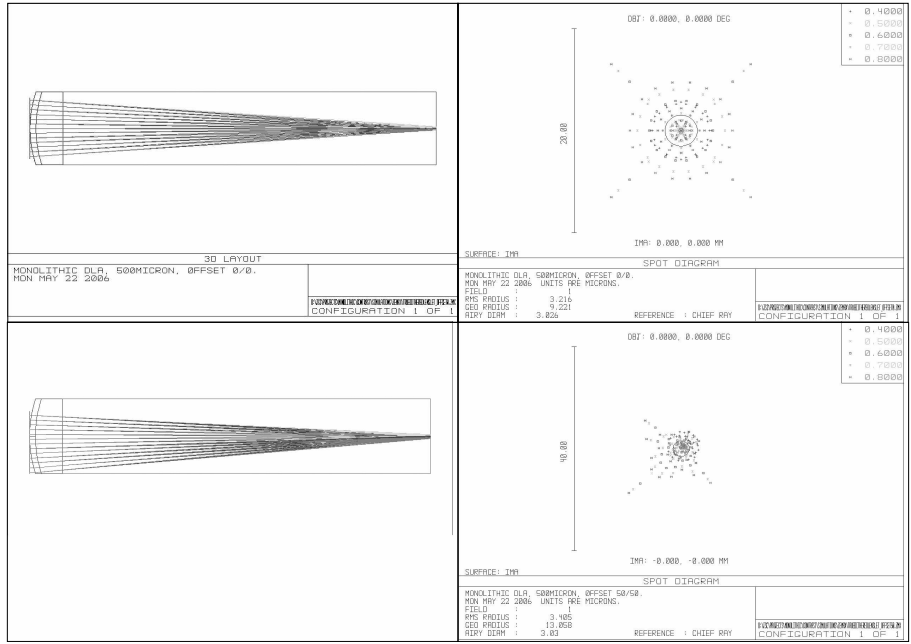}
   \end{tabular}
   \end{center}
   \caption{\label{fig:realditheredarray}
Simulation of monolithic dithered lenslet array elements. Above the concentric, below the 50 $\mu$m off-axis case.}
\end{figure}

As fig. \ref{fig:realditheredarray} shows, the quality of such a lenslet array made of PMMA on a BK7 substrate is near enough to the diffraction limit to be used to feed fibers, because the spot sizes are small in comparision to fiber cores being of typically 50 - 100 $\mu$m in diameter.

\begin{figure}[H]
   \begin{center}
   \begin{tabular}{c}
   \includegraphics[width=10cm]{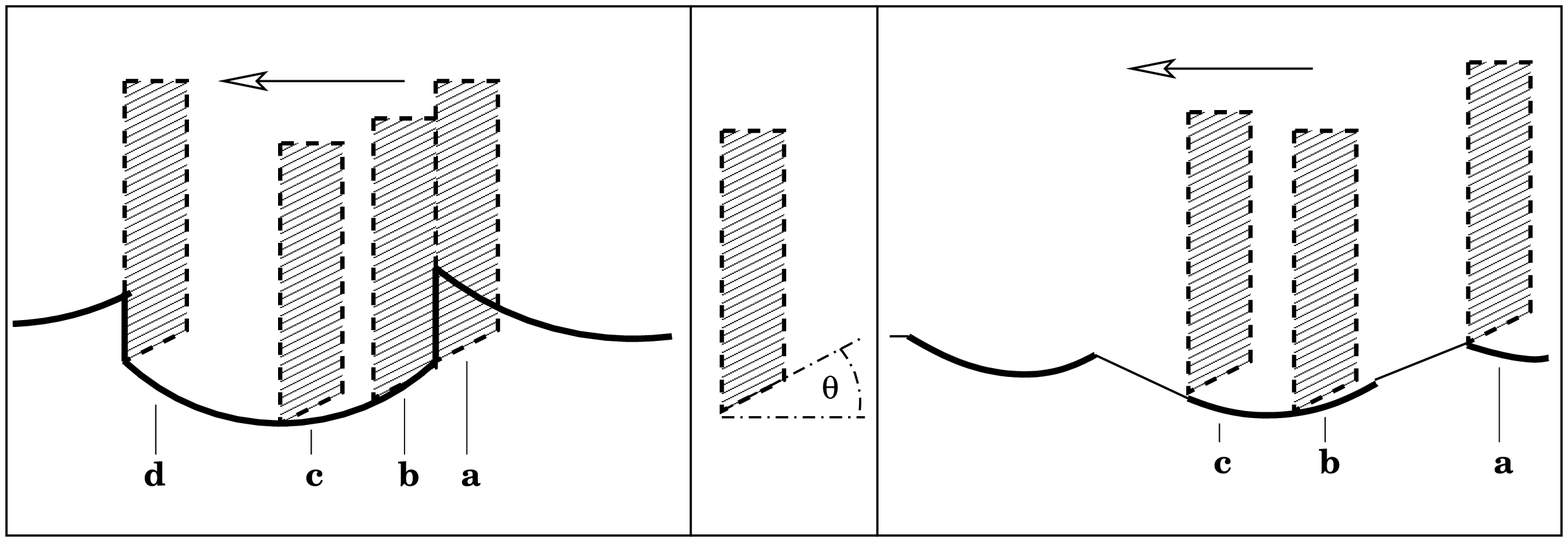}
   \end{tabular}
   \end{center}
   \caption{ \label{fig:cliffs} 
The cliff problem. Left dense array with discontinuities, right sparse array with linearly connected element borders. Center: Definition of the clearance angle $\theta$.}
\end{figure}

While the ZEMAX simulation as seen in fig.\ref{fig:realditheredarray} indicates a working system, difficulties arise in the context of manufacture of the negative preform necessary to press the PMMA material into shape on the BK7 substrate. Assessing the feasibility, the result was that neither slow-tool-servo, nor ruling or other milling techniques are able to produce such a form, the reason being the discontinuity between adjacent lenslets caused by the dithering pattern. While a concentric lenslet array has the adjacent lenslets touching each other on the same level, in the dithered case cliff-like structures occur that cannot be machined with a finite tool size. As fig. \ref{fig:cliffs}, left side, shows, there are several collisions of the tool and the workpiece. In (a), the tool would cut off a major part of the workpiece. Position (b) can be achieved, but a dead zone having the same width than the tool would be bare. While (c) works, at (d) there would be either a stop of the feed, or a infinetely fast vertical tool motion necessary. Furthermore there would be no space for the chips which are torn off by the tool tip. The collisions are caused by the finite thickness of the tool and the small clearance angle $\theta$ as depicted in fig \ref{fig:cliffs} necessary to hold the tool safely and vibration free. In a sparse array as shown to the right, the height differences can be bridged by a tool path that can be linear or of any shape until the surface tilt reaches the clearance angle.

\subsubsection{Sparse lenslet array approach}

\begin{figure}[H]
   \begin{center}
   \begin{tabular}{c}
   \includegraphics[width=6cm]{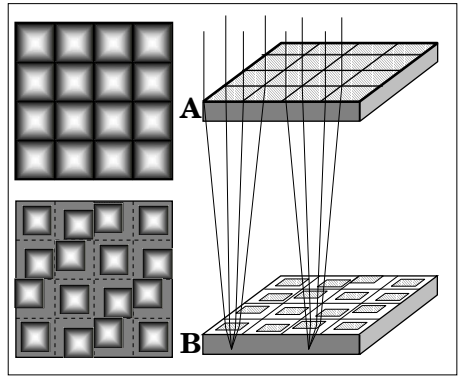}
   \end{tabular}
   \end{center}
   \caption{ \label{fig:sparselensarray_principle} 
2-element lens dithering approach. A: Dense concentric lens array. B: Sparse, dithered lens array.}
\end{figure}

\begin{figure}[H]
    \begin{center}
   \begin{tabular}{c}
   \includegraphics[width=12cm]{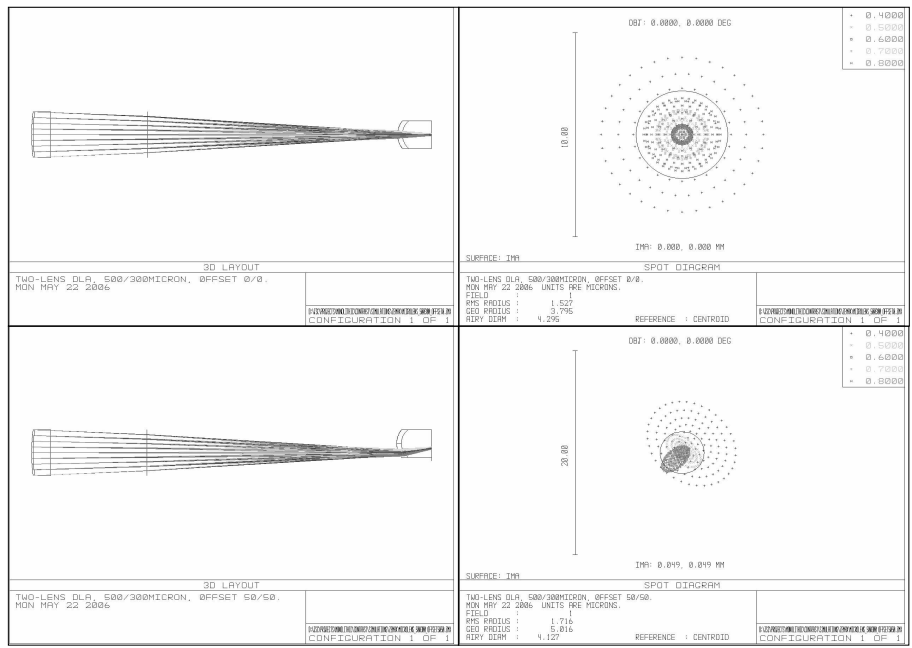}
   \end{tabular}
   \end{center}
   \caption{ \label{fig:sparselensarray} 
Simulation of 2-element dithering. A 500 $\mu$m classical lenslet array is followed by a sparsely populated area of 300 $\mu$m lenslets. While these lenslets are of classic shape, their offsets generate the dithering effect. The f/100 input beam is directed off by 50 $\mu$m, but telecentricity is lost and the 2nd lenslet curvature is very steep.
}
\end{figure}

A first idea to circumvent the problem with the discontinuities was a two-lens approach as seen in fig.\ref{fig:sparselensarray_principle}. The light enters a conventional lenslet array of relatively long focal length, filling the full aperture. Behind this array, another sparsely populated array with excentric lenses performes the dithering. While a simulation shows that the image quality is close to the diffraction limit (see fig. \ref{fig:sparselensarray}), the final focus displacement is difficult to achieve. Apart from a very short focal length for the 2nd lenslet array, the telecentricity is obviously lost. Such a design would be difficult to produce, and not very useful in practice.

\subsubsection{Micro-prism array approach}

\begin{figure}[H]
   \begin{center}
   \begin{tabular}{c}
   \includegraphics[width=6cm]{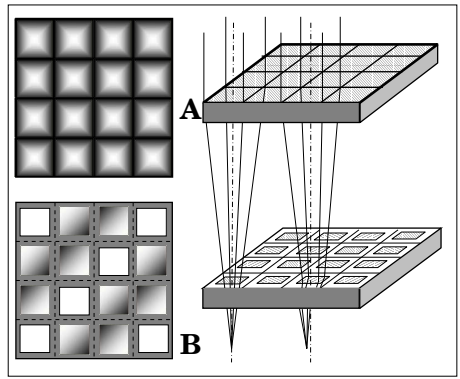}
   \end{tabular}
   \end{center}
   \caption{ \label{fig:sparseprismarray_principle} 
Dithering by use of a sparsely populated double micro-prism array (B) behind a dense, concentric lens array (A). The prism tilt is not shown in this illustration.}
\end{figure}

\begin{figure}[H]
   \begin{center}
   \begin{tabular}{c}
   \includegraphics[width=12cm]{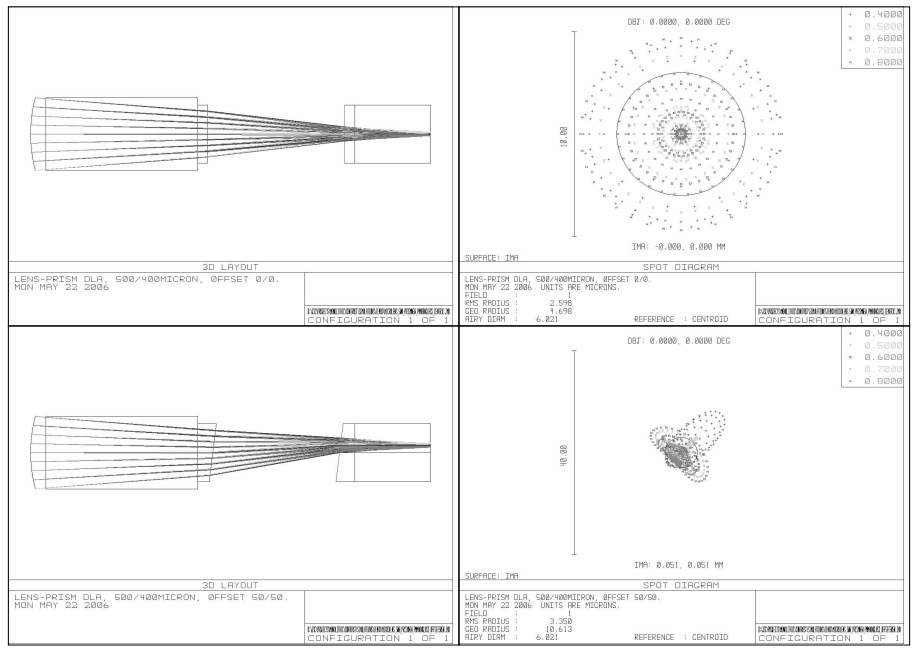}
   \end{tabular}
   \end{center}
   \caption{ \label{fig:prismsolution} 
Lenslet + prism array solution, example of 500 $\mu$m lenslets combined with 400 $\mu$m prisms, PMMA on BK7 substrate. Top to bottom: Offsets of 0 and 50 $\mu$m, left column layout, right column the corresponding spot diagrams.}
\end{figure}

To achieve telecentricity, another approach was tested at the expense of a third optical element per aperture. As fig. \ref{fig:sparseprismarray_principle} shows, the sparse lenslet array is replaced by a sparse prism array. Each micro-prism either just transmits the light (flat prism), or bends it before it passes the substrate of the 2nd array. In both cases, a corresponding prism of opposite tilt is located on the back side of the subtrate, regaining telecentricity. The simulation in fig. \ref{fig:prismsolution} shows that displacements of the desired 50 $\mu$m are possible without too much of image deterioration, taking into account that fiber core diameters are much larger than the aberrations shown.

\clearpage

\section{Conclusions}

The theoretical work has shown that free-form machined arrays may be used to tackle the problem of lenslet crosstalk due to spikes. However, the slow-tool servo is not a feasible method for creation of a monolithic dithered lenslet array. The design had to be adapted to the manufacturing process, on the expense of two more elements and one additional substrate. While the micro-prism array may have some potential for general situations, where beams have to be translated depending on the element they are entering, this is a tradeoff between contrast gain on one, but complexity and scatter on the other side. It is planned to test the manufacture of a dithered array using the slow-tool servo method and to compare it with a monolithic array that can be produced by use of a diamond-machined composite of linear arrangements. The comparision of both techniques will enable a final judgement if the slow-tool servo method is the preferable way of manufacture for dithered lenslet array masters.




\bibliography{references}   
\bibliographystyle{spiebib}   

\end{document}